# Active Latitude Oscillations Observed on the Sun

A. Kilcik[1], V. Yurchyshyn[2,3], F. Clette[4], A. Ozguc[5], J.-P. Rozelot[6]

[1]Akdeniz University Faculty of Science, Department of Space Science and Technologies, 07058, Antalya, Turkey

[2]Big Bear Solar Observatory of New Jersey Institute of Technology, Big Bear City, CA 92314, USA

[3]Korea Astronomy and Space Science Institute, 776 Daedeok-daero,, Daejeon, 305-348, South Korea

[4]Royal Observatory of Belgium Av. Circulaire, 3 - B-1180 Brussels, Belgium

[5]Kandilli Observatory and Earthquake Research Institute, Bogazici University, 34684 Istanbul, Turkey

[6]Nice University (OCA), Bvd. de l'Observatoire, CS 34229, F-06304 Nice Cedex 4, France

**Abstract**

We investigate periodicities in mean heliographic latitudes of sunspot groups, called active latitudes, for the last six complete solar cycles (1945-2008). For this purpose, the Multi Taper Method and Morlet Wavelet analysis methods were used. We found the following: 1) Solar rotation periodicities (26-38 days) are present in active latitudes of both hemispheres for all the investigated cycles (18 to 23). 2) Both in the northern and southern hemispheres, active latitudes drifted towards the equator starting from the beginning to the end of each cycle by following an oscillating path. These motions are well described by a second order polynomial. 3) There are no meaningful periods between 55 and about 300 days in either hemisphere for all cycles. 4) A 300 to 370 day periodicity appears in both hemispheres for Cycle 23, in the northern hemisphere for Cycle 20, and in the southern hemisphere for Cycle 18.

# 1. Introduction

The Sun is the only star that provides us with a possibility to perform very high resolution detailed observations of variations occurring on its surface and in higher atmospheric layers by using variety of special techniques and observational apparatus. These variations may include changes in the number and size of sunspots, which are called solar activity variations. They can be investigated through various indicators such as the sunspot number (SSN), sunspot area (SSA), the 10.7 cm solar radio flux (F10.7), total solar irradiance (TSI), etc. All these indicators are closely related to each other. Due to its very long temporal coverage (∼400 years), the SSN is the most widely used index of solar activity.

The above solar activity indicators generally show cyclic behavior with periods ranging from days to thousands of years. The best known solar cyclic variations are the eleven year sunspot cycle that was first discovered by Heinrich Schwabe in 1843 using 17 years of sunspot observations (Schwabe, 1844), and the 27 day solar rotation periodicity that is induced by large long-lived active regions (ARs) with lifetimes longer than one solar rotation. Investigations of possible periodicities other than 11 years and 27 days have been of interest for a long time. Many researchers have investigated these periodicities using various solar activity indicators and methods. They found various periods between 512 and 25 days: such as 450-512 days, 280-364 days, 210-240 days, 150-170 days, 120-130 days, 110-115 days, 73-78 days, 62-68 days, 51-58 days, 41-47 days, and 25-37 days (Lean and Brueckner, 1989; Bai and Sturrock, 1991; Rozelot, 1993; Krivova and Solanki, 2002; Bai, 2003; Ozguc, Atac, and Rybak, 2004; Kane, 2005; Knaack, Stenflo, and Berdyugina, 2005; Obridko and Shelting, 2007; Lara et al., 2008; Kilcik et al., 2010, 2014; Chowdhury and Dwivedi, 2011; Scafetta and Willson, 2013; Choudhary et al., 2014; Ravindra and Javaraiah, 2015 and references therein).

However, there are inconsistencies between the periods determined by different studies. These inconsistencies possibly come from the analyzed data, the methods used or the investigated time intervals. Moreover, different solar activity indicators have different origins depending on the layer of the solar atmosphere where they are measured (Bouwer, 1992).

At the start of a sunspot cycle, sunspot groups tend to appear around ± 40° heliographic latitude on the Sun. As the cycle progresses, the heliographic latitude of sunspots groups, hereafter called "active latitude", decreases gradually down to a few degrees and then, while the old sunspot cycle fades away, sunspot groups of the new cycle start appearing again at high latitudes. This gradual equatorward drift of sunspot groups has been well known for a long time (Maunder, 1904; Howard and Gilman 1986; Lustig and Wöhl, 1994 and references therein).

Here, we searched for periodic behavior in the daily mean active latitudes. We report that some of the above-mentioned published periodicities found in solar rotation are also found in active latitude oscillations. Accurate determination of these oscillations may provide a clue for improving our understanding of the underlying mechanisms responsible for this multi-periodic behavior of solar activity.

In Section 2, we describe the methodology and the data analysis. In Section 3, we present the results and in Section 4, we draw conclusions and briefly discuss their implications.

**2. Data and Methods**

The Royal Greenwich Observatory and the USAF/NOAA sunspot group data used in this study were taken from the National Aeronautics and Space Administration (NASA) data service center[1]. The data include the date (year, month, and day), the NOAA active region number, the heliographic coordinates (latitude and longitude) of sunspot groups, *etc.* We used the heliographic latitudes of all sunspot groups measured over the period from 1945 to 2008. For each day, we calculated arithmetic average of the latitudes of all observed groups to produce what we call the daily active latitude. We limited our analysis to Solar Cycles 18-23 due to the abrupt change in the ratio of the umbral areas to total spot area occurring in 1941/1942 (see NASA data service center). As these changes may affect the distribution of active latitudes, we chose to avoid this transition. To eliminate the overlap between successive solar cycles, we did not use data near solar cycle minima, namely in the interval from one year before to one year after a minimum year. The hemispheric active latitude data were produced for each solar cycle and for each hemisphere separately. In order to improve the detection of long term oscillatory patterns in the daily active latitude data, we first smoothed all the data sets by a simple 100-point (100-day) boxcar running mean.

---
1 http://solarscience.msfc.nasa.gov/greenwch.shtml

To further reduce possible contamination of the results by random noise in the data, we calculated the standard deviation of daily values relative to the global average value of each solar cycle. Then we derived the best second order polynomial fit to the data for each cycle, from Cycle 18 through 23. Based on these standard deviations and fitted values, we derived the upper and lower significance limits, set at two standard deviations, for each cycle separately. All data points lying outside the significance limits were dropped before applying the period analysis. Due to the existence of various data gaps over the whole time interval, we worked on data segments, spanning each a four-year interval centered on each solar maximum (1946-1949, 1957-1960, 1967-1970, 1979-1982, 1989-1992, and 1999-2002) in our period analysis. Those time segments still include very small (a few days) time interval gaps, which are due to spotless days, noise reduction procedure, and lack of observation, compared to cyclic data. These short gaps were filled by cubic spline interpolation and as a final procedure, we applied a three-step boxcar running average for smoothing each resulting uninterrupted data segment.

We applied two analysis methods to the data resulting from the above pre-processing: the Multi Taper Method (MTM) and the Morlet wavelet analysis. The MTM allows us to infer possible periodicities having a high overall degree of statistical significance over the full duration of each investigated time segment. The method uses orthogonal windows (or tapers) to obtain the estimate of the power spectrum (for more details see Thomson 1982; Ghil et al. 2002). This method has been extensively utilized in solar physics (Kilcik et al. 2010, 2014; Mufti and Shah 2011, and references therein). Here we used three sinusoidal tapers, and the frequency range was limited to 0.04 day$^{-1}$ (25 days). The significance test was carried out assuming that the noise has a red spectrum, since the data have a higher power density at low frequencies and a lower power density at high frequencies. We considered that a signal was detected when the 95 % confidence level was reached.

The wavelet analysis has an advantage compared to other period analysis methods, in the sense that it allows evaluating possible temporal variations of the oscillations detected in the active latitudes by the MTM analysis. This method has also been extensively used in solar physics (e.g. Torrence and Compo, 1998; Krivova and Solanki, 2002; Deng et al., 2013; Choudhary et al., 2014; Kilcik et al. 2014; Singh and Badruddin, 2014 and references therein). We used the Morlet wavelet analysis package from IDL (Interactive Data Language). The resulting scalograms were used to study both the existence and the evolution of periodicities. Note that for the Morlet wavelet analysis, the significance test was also carried out assuming a red-noise spectrum.

## 3. Results

Figure 1 shows the temporal evolution of the active latitudes, separately for the northern and southern hemispheres and for each solar cycle (Cycles 18 through 23).

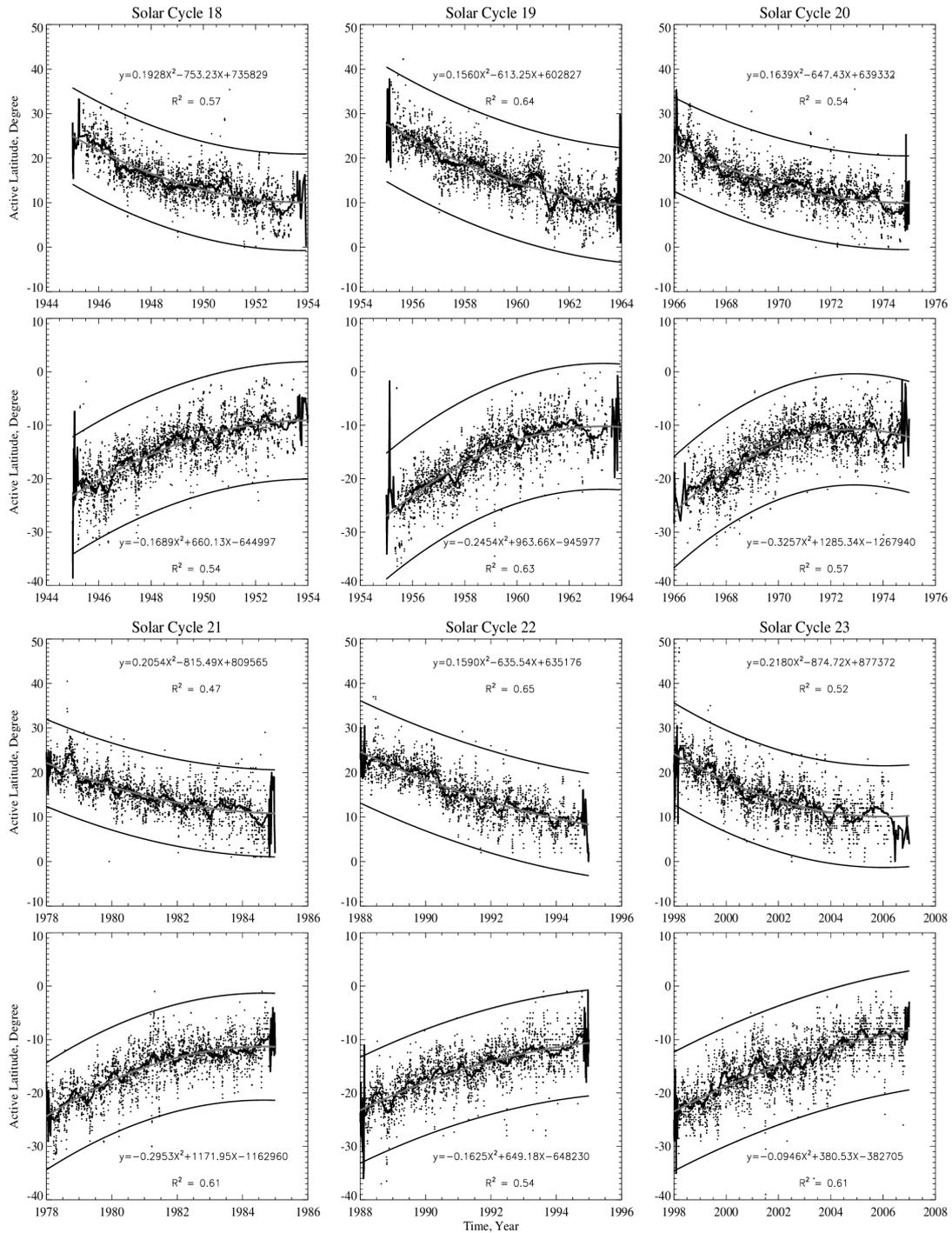

***Figure 1.*** *Temporal variations of active latitudes for the northern and southern hemispheres plotted separately for each cycle (Cycles 18 to 23). The dots represent the daily average heliographic latitude, the black solid line is the 100-days running mean, and the gray solid*

*line is the best second order polynomial fit. Solid curves which are parallel to the polynomial curve indicate the 2σ limits (σ = standard deviation). Large oscillations seen at the edges of the plotted cycles are due to the 100-steps running mean smooth procedure.*

Figure 1 highlights two interesting tendencies. First, the active latitudes display a steady drift towards the equator during the entire length of the solar cycle, which reflects the well-known tendency for ARs to appear closer to the solar equator as the solar cycle progresses (Butterfly diagram). This drift can be closely approximated by a second order polynomial (see the smoothed curve). Secondly, clear oscillations of the active latitudes occur around the main monotonous equatorward drift in all the investigated cycles.

In order to reveal the details of these periodic variations, we plot in Figures 2 to 7 the MTM spectrum and the Morlet wavelet scalograms for the maximum phase of each cycle.

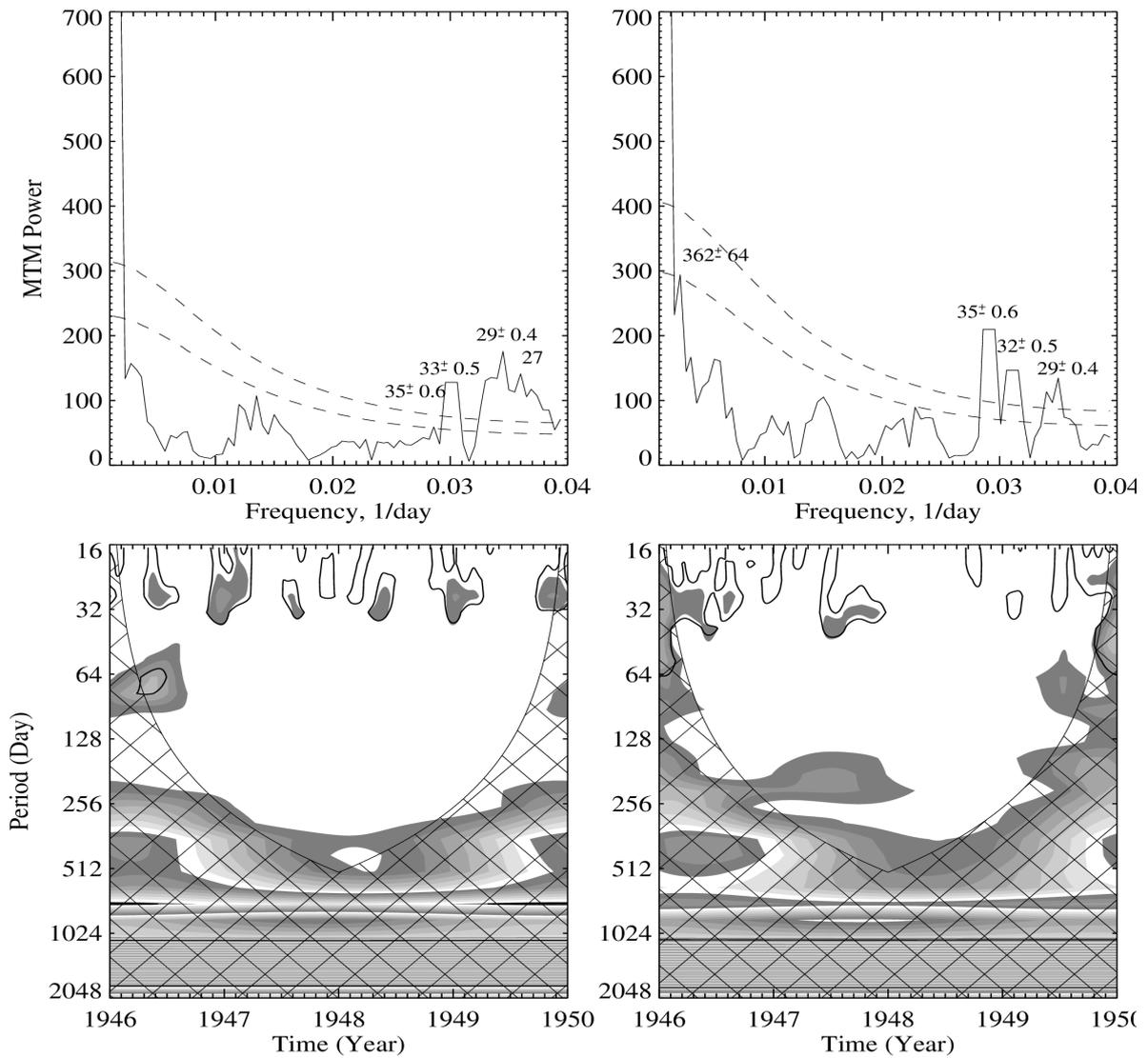

*Figure 2. Results of multi-taper method (MTM, upper panels) and Morlet wavelet analysis (lower panels) of active latitudes for the northern (left panels) and southern hemispheres (right panels) near the maximum of solar Cycle 18 (1946-1949). The black contours in the wavelet scalogram indicate a 95% confidence level and the hatched area below the thin black line is the cone of influence (COI). In the upper panels, the main peaks are labeled with the period in days. The horizontal dashed lines indicate the 95% and 99% confidence levels of the MTM power spectrum.*

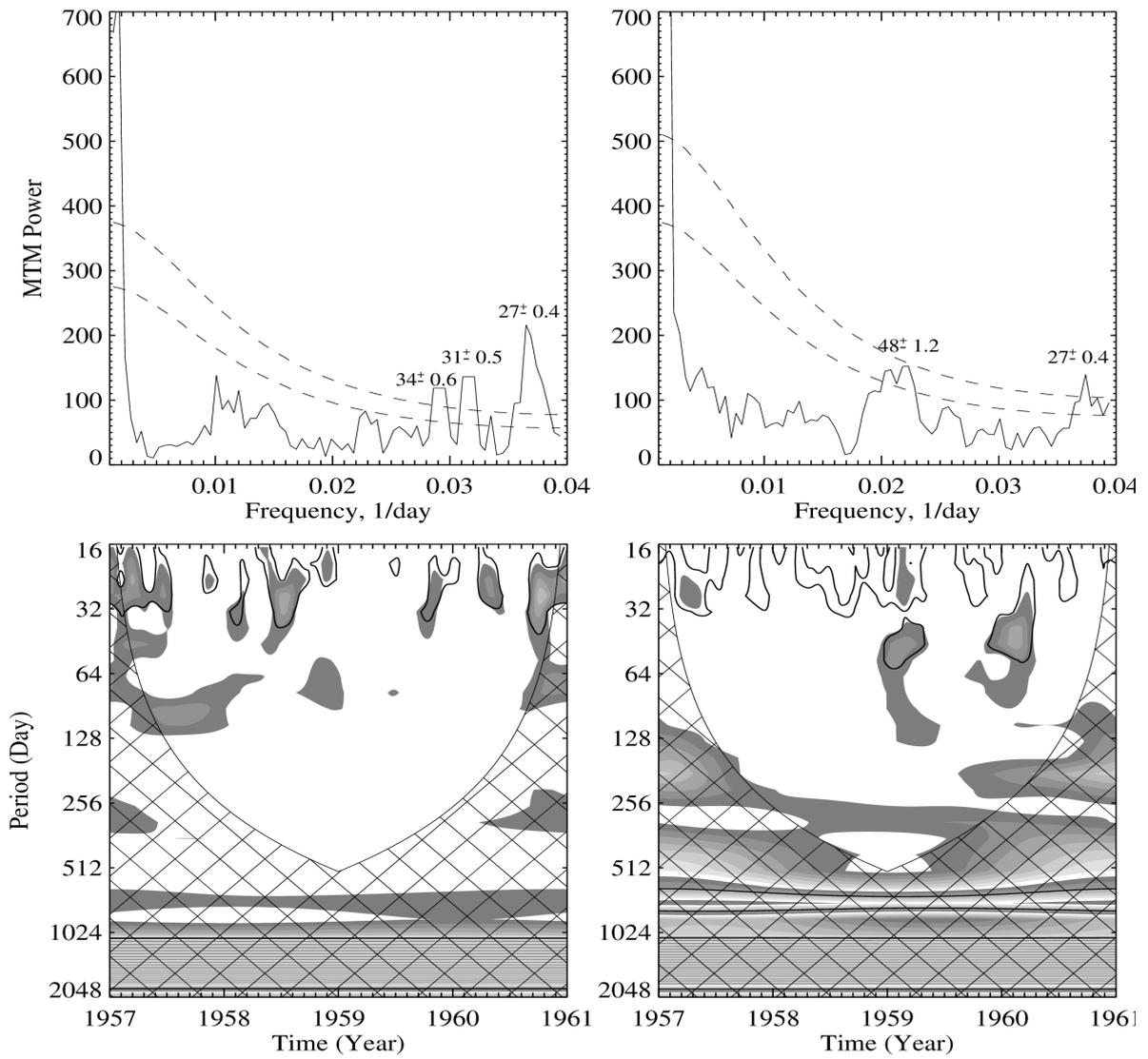

*Figure 3. Same plot as in Figure 2 for Cycle 19 (1957-1960).*

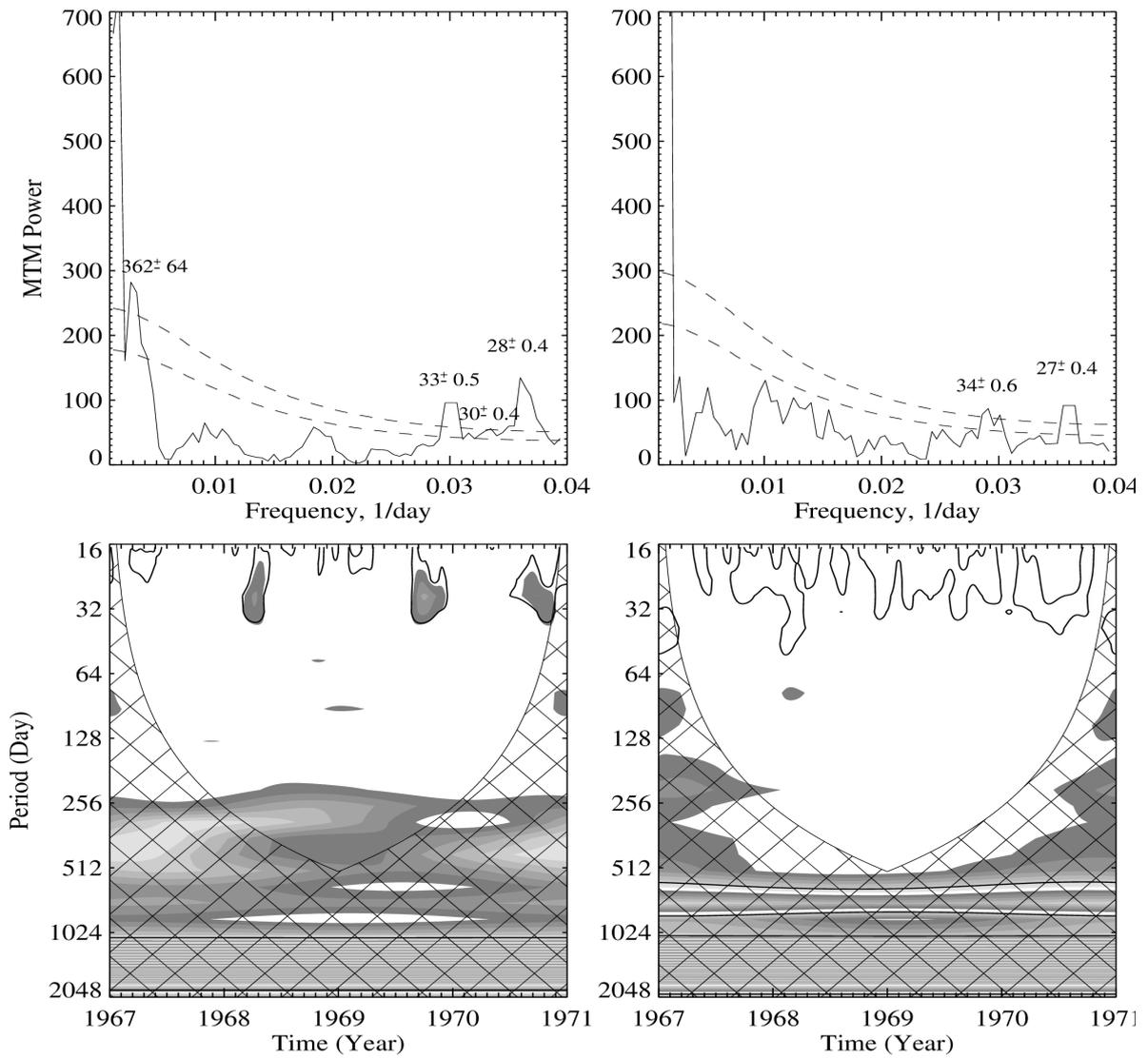

*Figure 4.* Same plot as in Figure 2 for Cycle 20 (1967-1970).

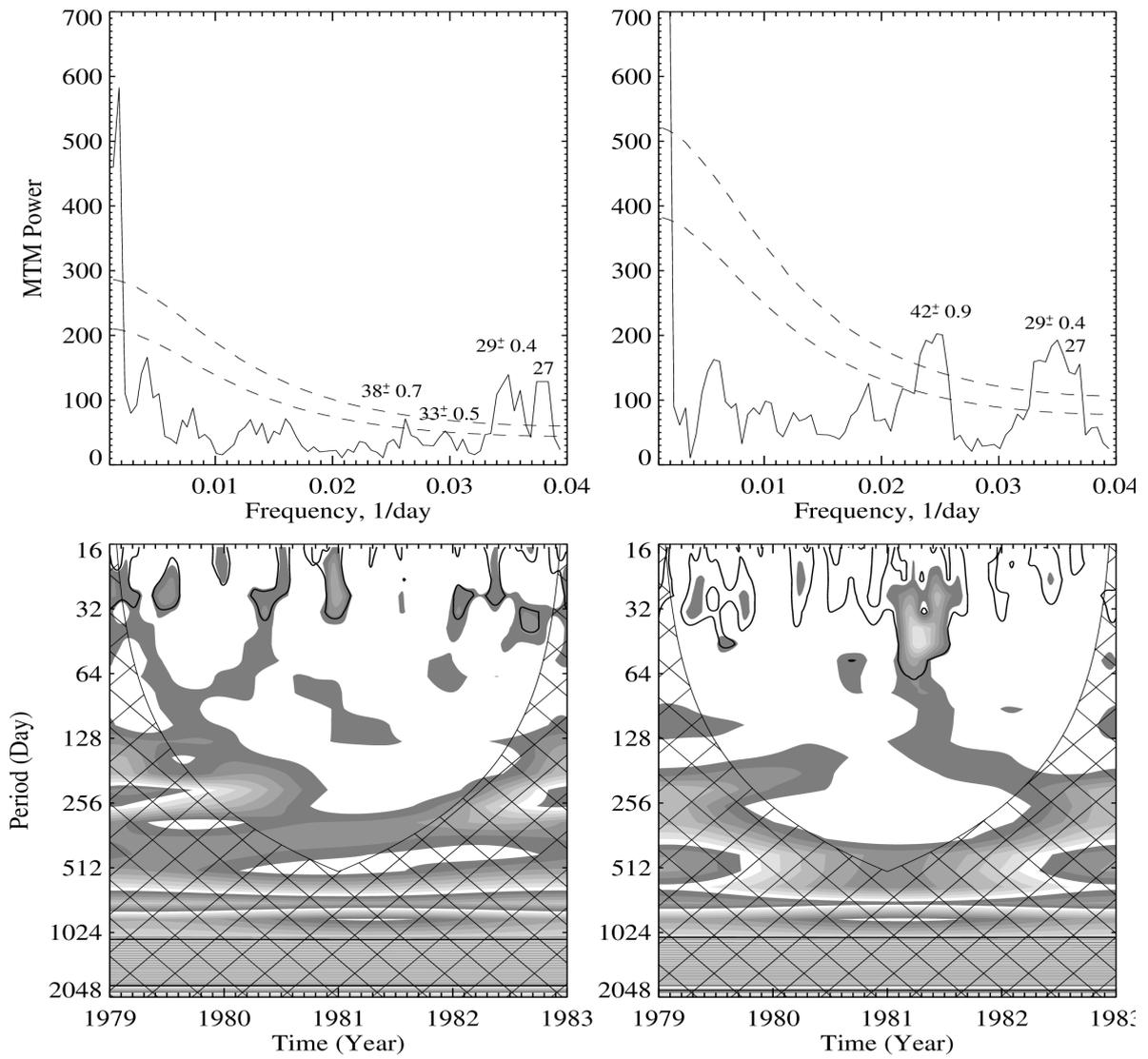

*Figure 5. Same plot as in Figure 2 for Cycle 21 (1979-1982).*

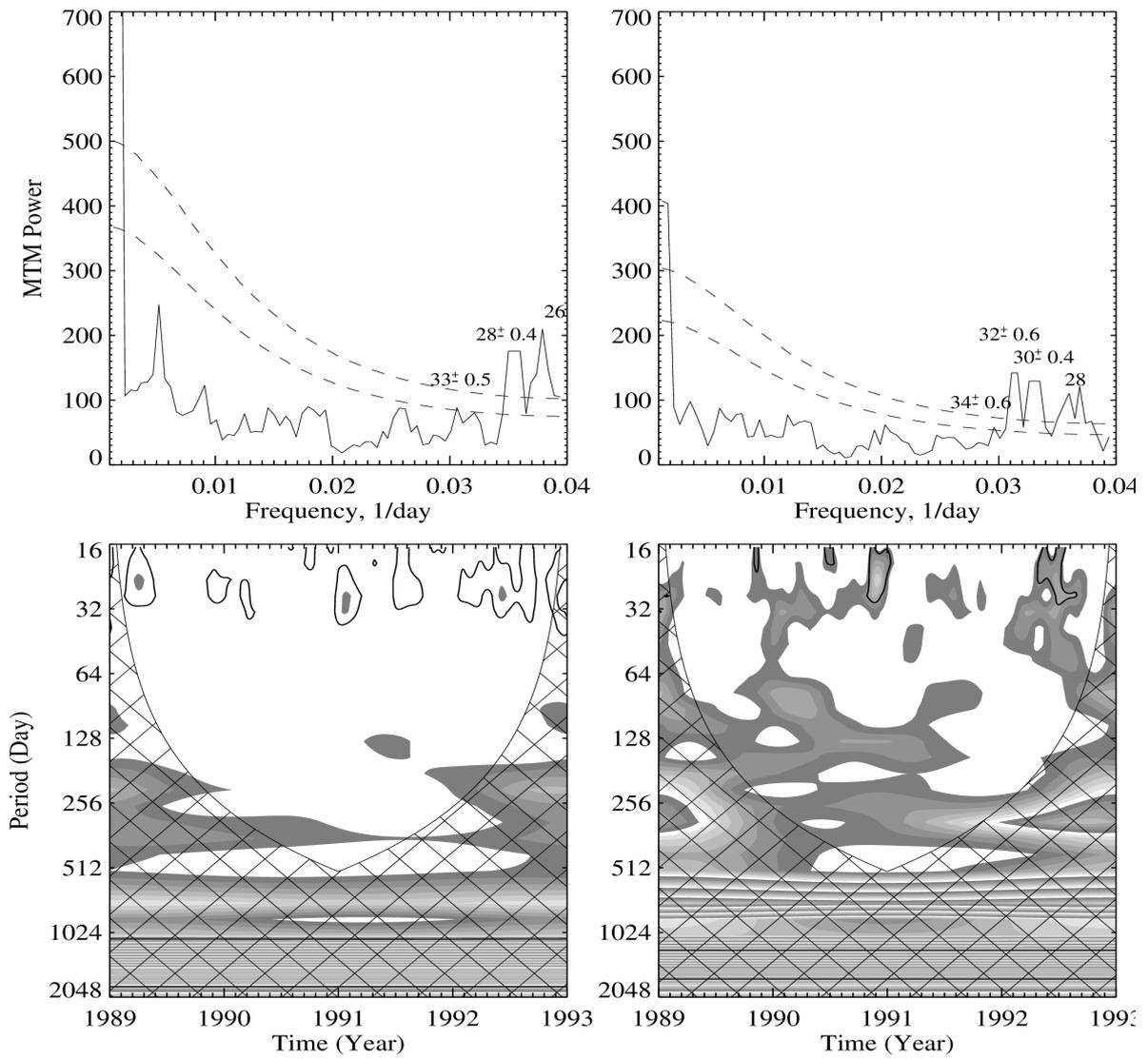

*Figure 6.* Same plot as in Figure 2 for Cycle 22 (1989-1992).

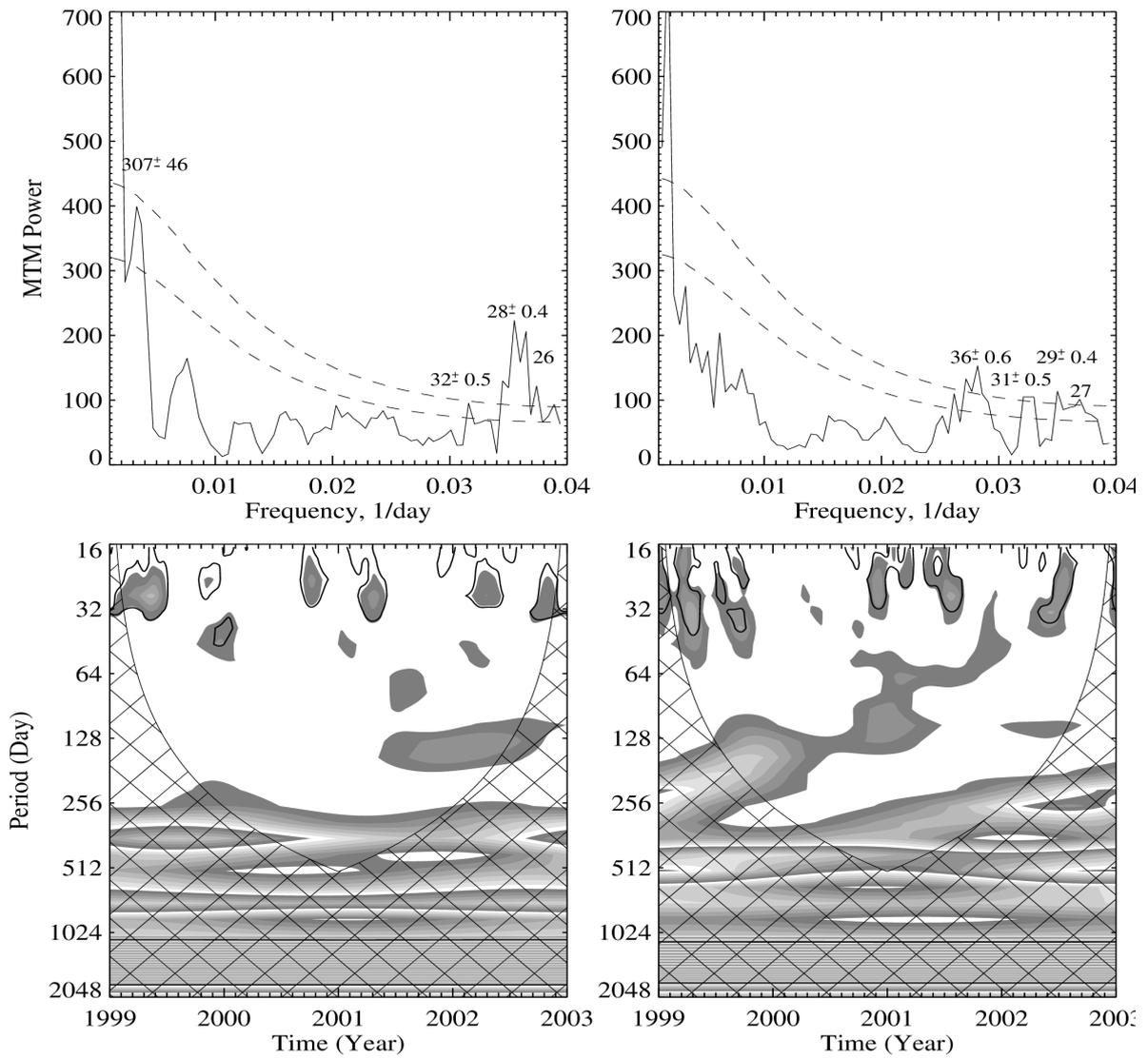

*Figure 7. Same plot as in Figure 2 for Cycle 23 (1999-2002).*

| Cycle No | Period (Day) | | |
|---|---|---|---|
| | 26 - 38 | 40 - 55 | 300 - 370 |
| Cycle 18 North | + > 95 | - | - |
| Cycle 18 South | + > 95 | + < 95 | + > 95 (362) |
| Cycle 19 North | + > 95 | + < 95 | - |
| Cycle 19 South | + > 95 | + > 95 | - |
| Cycle 20 North | + > 95 | + < 95 | + > 99 (362) |
| Cycle 20 South | + > 95 | + < 95 | - |
| Cycle 21 North | + > 95 | - | - |
| Cycle 21 South | + > 95 | + > 95 | - |
| Cycle 22 North | + > 95 | - | - |
| Cycle 22 South | + > 95 | - | - |
| Cycle 23 North | + > 95 | + < 95 | + > 95 (307) |
| Cycle 23 South | + > 95 | + < 95 | + < 95 (307) |

*Table 1.* *MTM periods obtained for different solar cycles and hemispheres. The first column lists the periods detected in this study while the other columns show the presence of given periodicity in the northern and southern solar hemispheres for each solar cycle.*

Based on the above figures and the table, we observe that the solar rotation period (26 – 38 days) is present in the active latitudes data for both hemispheres and in all investigated solar cycles. On the other hand, we do not find any meaningful period in the range of 55-300 days, both in the MTM spectra and Morlet wavelet scalograms of all data sets. A 50 day (40-55 days) period is detected with a confidence level exceeding 90% in all the investigated cycles, except for Cycle 22 and for the northern hemispheres of Cycle 18 and 21, where it remains below the 90% confidence level. These periods are also seen in the wavelet scalograms of the same cycles. A 300-370-day periodicity appears in both hemispheres in Cycle 23, peaking at a 307 day period, and only in the southern hemisphere of Cycle 18 and the northern hemisphere of Cycle 20, then peaking at a 362 day period.

**Discussion and Conclusions**

In this study, we investigated periodic variations of the daily active latitude of sunspot groups over the time interval 1945 to 2008. We report the following results:

1. The time profiles of the daily active latitudes both in the northern and southern hemispheres show periodic variations superimposed on the main equatorward drift marking the entire duration of all investigated solar cycles (Cycles 18 through 23).
2. Solar rotation periodicities (26-38 days) are present in the active latitude data for both hemispheres during all the investigated cycles without any hemispheric preference.
3. A broad 50-day period (spanning the 40-55 days interval) is found in all the investigated cycles with at least a 90 % confidence level, except Cycle 22 and in the northern hemisphere of Cycles 18 and 21, where it was detected with a much lower confidence level below 90%.
4. There are no meaningful periodicities between 55 and about 300 days in both hemispheres for all cycles.
5. A 300-370 day periodicity appears only in Cycles 18, 20 and 23.

We analyzed the daily averaged active latitudes of solar AR and found that the equatorward shift of the sunspot group is not fully monotonous and also shows cyclic fluctuations. We find that previously reported periods in other solar activity indices do also appear in the daily mean active latitudes: namely the SSN (Yin et al., 2007; Chowdhury and Dwivedi, 2011; Kilcik et al., 2014; Choudhary et al, 2014 and reference therein), and SSA (Chowdhury, Khan, and Ray, 2009; Choudhary et al, 2014 and reference therein). As such periods are common with multiple global solar activity tracers (SSN, SSA, F10.7), we can thus speculate that they form the surface manifestation of large-scale plasma flows at the solar surface or rooted below it, deeper in the convective region.

Still, we must point out that for the same solar cycles analyzed here, most authors reported periodicities between 54 and 300 days (Yin et al., 2007; Chowdhury and Dwivedi, 2011; Kilcik et al., 2014; Choudhary et al, 2014 and reference therein). By contrast, we did not find any meaningful period in this range in the active latitude data. This result may indicate that periods between 54 and 300 days may be specific only to the other activity indices. The 315 to 348-day periodicities were also previously detected in some solar activity indices (Lean and Brueckner, 1989; Scafetta and Willson, 2013; Kilcik et al., 2014 and reference therein). Kilcik et al (2014) analyzed sunspot counts in four categories and found this periodicity in the counts of large and well-developed sunspot groups with more than 95% significance level, while no periodicities were detected in the counts of simple A- and B-type sunspot groups. Lean and Brueckner, (1989) analyzed the 10.7 cm solar radio flux, the SSN, the plage index and the sunspot blocking function for solar Cycles 19, 20 and 21. They found that a peak near 323 days in peridograms of all those data series. Thus, they concluded that these periods may have a real solar origin. Later, Beck and Giles (2005) analyzed measurements of meridional flows obtained from the Michelson Doppler Imager (MDI) from 1996 May through 2001 July, and found a sinusoidal variation with a period of one year due to an error in the determination of the solar rotation axis. Here, we analyzed much longer data set, (from 1945 through 2008) and found the same period in the active latitude data in Cycles 18, 20, and 23 with at least 95% confidence level. Large-amplitude variations are seen clearly in Figure 1 for all active latitude data sets. We also detect distinct periodicities (307 and 362 days) for different cycles. Thus, we speculate that this period is not due to an error in the determination of the rotation axis and that it has a real solar origin as suggested by Lean and Bruckner (1989). Moreover, we could not find any harmonics (180 days, 90 days etc.) of such a periodicity, which further supports the above interpretation.

All these results may indicate that solar activity indices and the underlying active latitudes oscillate in the same way. In other words, the strength of solar activity and its spatial distribution show the same periodic patterns. This suggests the existence of a direct connection between the efficiency of the sub-surface dynamo (translated by the values of activity indices and solar flux measurements) and its spatial localization in the course of a solar cycle (latitudinal and longitudinal distribution of active regions). Finally, we observe that the periodicities found here are specific to each solar cycle and to each hemisphere. Only the solar rotation periodicity (26-38 days) does not show any difference between hemispheres, which can be expected as solar rotation is a permanent global property largely independent from the solar cycle and dynamo processes.

The present results should now be combined in more detail with equivalent variations of other solar activity indices (SSNs, SSAs, etc.), for instance, by determining the phase coherence between corresponding periods shared by several data series and the correspondence with the distribution of active regions in longitude (active longitudes). This future work could help deciphering how global temporal and spatial properties of active regions and the underlying solar dynamo may control the irregular evolution of solar activity from cycle to cycle.

**Acknowledgement** The sunspot group data used in this study was retrieved from the National Aeronautics and Space Administration (NASA) on-line data archives. The wavelet software was provided by C. Torrence and G. Compo, and is available at http://paos.colorado.edu/research/wavelets/. This study was supported by the Scientific and Technical Council of Turkey by the Project of 115F031. V. Yurchyshyn acknowledges support from AFOSR Award FA9550-15-1-0322 and NSF AGS-1146896 grant and Korea Astronomy and Space Science Institute. F. Clette would like to acknowledge financial support from the Belgian Solar-Terrestrial Center of Excellence (STCE; http://www.stce.be). J.-P. Rozelot would like to thank the International Space Science Institute in Bern (CH) for its support of this study.